\documentclass
[prl,twocolumn,superscriptaddress,floatfix,10pt]{revtex4}%
\usepackage{amsfonts}
\usepackage{amsmath}
\usepackage{amssymb}
\usepackage{graphicx}%
\newcommand{\LL }{\mathcal{L}}

\newcommand{\ee }{\ell }

\begin{document}
\title{Causal Self-Dual Electrodynamics}

\author{Jorge G. Russo}
\affiliation{ Instituci\'o Catalana de Recerca i Estudis Avan\c{c}ats (ICREA),\\
Pg. Lluis Companys, 23, 08010 Barcelona, Spain, {\sl and} \\
Departament de F\' \i sica Cu\' antica i Astrof\'\i sica and Institut de Ci\`encies del Cosmos,\\ 
Universitat de Barcelona, Mart\'i Franqu\`es, 1, 08028
Barcelona, Spain.}
\author{Paul K. Townsend}
\affiliation{Department of Applied Mathematics and Theoretical Physics, Centre for
Mathematical Sciences, University of Cambridge, Wilberforce Road, Cambridge,
CB3 0WA, UK.}

\begin{titlepage}
\vfill
\end{titlepage}

\begin{abstract}

Many theories of nonlinear electrodynamics (NLED) that have been proposed in physical contexts involving strong fields are causal for weak fields but acausal for strong fields. We show  that for any such theory there is a unique causal and self-dual (electromagnetic duality invariant) theory with the same Lagrangian at zero magnetic field.
This follows from a construction of the general causal self-dual NLED, which shows that strong-field causality is implied by weak-field causality for self-dual theories.  
We illustrate our results with explicit examples.



\end{abstract}

\maketitle

\setcounter{equation}{0}

Theories of nonlinear electrodynamics (NLED) are interacting generalisations of Maxwell 
electrodynamics that are traditionally defined (since the work of Born \cite{Born:1933qff}, Born and Infeld 
\cite{Born:1934gh}, and later Boillat \cite{Boillat:1966eyw} and Plebanski \cite{Plebanski:1970zz}) by a Lagrangian-density 
function $\mathcal{L}(S,P)$ of the two Lorentz scalars quadratic in the electric and magnetic fields:
\begin{equation}
S= \tfrac12(|{\bf E}|^2 -|{\bf B}|^2)\, , \qquad P= {\bf E}\cdot{\bf B}\, . 
\end{equation}
For the linear case, $\mathcal{L}\propto S$, since $P$ is a total derivative, 
and positive energy requires a positive proportionality constant. As the normalisation depends on a choice of units, we 
choose $\mathcal{L}=S$; this is the Maxwell case.  

An interacting example arises in the context of the effective action for QED at energy densities insufficient for electron-positron pair production.  The full effective action takes the form of an expansion in powers of $(S,P)$ {\sl and} their derivatives. As 
derivative terms generally introduce additional unphysical massive modes, they limit the applicability of effective field theory to 
energies below the scale that they introduce. However, the derivative terms are absent in the approximation of constant 
electric and magnetic fields, and the sum of the expansion in powers of $(S,P)$ yields the 1936 NLED of
Heisenberg and Euler \cite{Heisenberg:1936nmg}. Its validity, within the constant-field approximation, is not obviously limited to weak fields because NLED interactions do not introduce additional modes. For example, for specified wave-vector ${\bf k}$ there are still only two independent polarisations of any plane-wave perturbation of any constant uniform electromagnetic background. 

The similarity of the Heisenberg-Euler theory to the theories of Born and Infeld was noted at the time \cite{Dunne:2012vv}. However, Born's idea was that  Maxwell electrodynamics 
is already a weak-field approximation to an interacting relativistic {\sl classical} field theory with a maximal electric field strength determined by a new fundamental constant of nature, the ``Born constant''.  Born's motivations have since been justified, to some extent, by the appearance of the Born-Infeld (BI) theory as part of the effective worldvolume dynamics of the D3-brane of IIB superstring theory at weak string coupling; the 3-brane tension $T$ plays the role of the Born constant (see e.g. \cite{Tseytlin:1999dj}). A Born scale in classical electrodynamics is potentially relevant in contexts such as black holes \cite{Soleng:1995kn,Gunasekaran:2012dq}, magnetars \cite{Pereira:2018mnn,Kim:2022xum}, particle colliders \cite{Ellis:2022uxv,Price:2023cll}, and condensed matter via holography \cite{Jing:2010zp}; see e.g. \cite{Sorokin:2021tge,Russo:2024kto} for other references. 
%
An important consideration in many of these applications is causality. In applications to classical black holes physics, for example, causality is required for the existence of an event horizon.  

The necessary and sufficient conditions for $\mathcal{L}(S,P)$ to define a causal NLED were found by Schellstede et al. \cite{Schellstede:2016zue}, subject to an assumption that amounts to the existence of a weak-field limit. A pre-condition is 
\begin{equation}\label{pre-c}
\mathcal{L}_S >0\, , 
\end{equation}
where the subscript denotes a partial derivative; this is partly a sign convention. Then we have the conditions
\begin{equation}\label{Schell1}
\mathcal{L}_{SS}\ge 0\, , \qquad \mathcal{L}_{PP} \ge 0\, , \qquad 
\mathcal{L}_{SS}\mathcal{L}_{PP} - \mathcal{L}_{SP}^2 \ge0 \, ,  
\end{equation} 
which are also the conditions for convexity of $\mathcal{L}$, viewed as a function of the electric field ${\bf E}$ \cite{Bandos:2021rqy}. We interpreted them as weak-field causality conditions in \cite{Russo:2024kto} because causality violation for weak fields requires a negative birefringence index \cite{Bialynicki-Birula:1984daz,Russo:2022qvz}, which is excluded by convexity \cite{Bandos:2021rqy}. For brevity and clarity, we shall refer here to the inequalities of \eqref{Schell1} as the ``convexity conditions''.  

Finally, we have the condition \cite{Schellstede:2016zue}
\begin{equation}\label{Schell23}
\mathcal{L}_S > 2U \mathcal{L}_{SS}+ 2V\mathcal{L}_{PP} - 2P\mathcal{L}_{SP}\, ,  
\end{equation}
where
\begin{equation}
\label{UVdef}
U=\tfrac12(\sqrt{S^2+P^2}-S)\, ,\quad V=\tfrac12(\sqrt{S^2+P^2}+S)\, .
\end{equation}
Notice that this condition implies \eqref{pre-c} given \eqref{Schell1},  so \eqref{pre-c} is purely a convention for causal theories. A violation of \eqref{Schell23} requires strong fields. We have given an alternative derivation of this strong-field causality condition in \cite{Russo:2024kto}; we note here 
that it becomes the following much simpler condition when 
$\mathcal{L}$ is regarded as a function of $(U,V)$:
\begin{equation}\label{dos}
\mathcal{L}_U+2U \mathcal{L}_{UU}<0\ . 
\end{equation}

In NLED theories with only a few parameters, the convexity conditions \eqref{Schell1} usually reduce to a few sign choices, and most models considered in the literature are physical in this respect, i.e. causal for sufficiently weak fields. This is true of Born's original theory, Born-Infeld, and most other variants of Born's theory considered in applications mentioned above.  It has typically been assumed, implicitly, that this remains true for strong fields (as is the case for Born-Infeld \cite{Bialynicki-Birula:1984daz}). However, functions $\mathcal{L}(S,P)$ chosen simply because they appear promising for some  phenomenological purpose are unlikely to satisfy the strong-field 
causality condition \eqref{Schell23}; this was Born's approach and his original model is acausal \cite{Schellstede:2016zue}.  In fact, any NLED defined (like Born's original theory) by a Lagrangian density function $\mathcal{L}(S)$ (i.e. no $P$-dependence) is acausal \cite{Schellstede:2016zue}. Many others more similar to the Born-Infeld theory are also acausal \cite{Russo:2024kto}. 

The Heisenberg-Euler function $\mathcal{L}(S,P)$ is known only in an implicit form that determines
the expansion to any order in powers of $(S,P)$. The truncation to quadratic order is acausal for strong fields but the truncated-expansion approximation is then invalid \cite{Schellstede:2016zue}. It is unknown whether the un-truncated theory is causal; it might be expected to violate causality for electric fields strong enough for pair production, but then it should be replaced by QED. However, a similar escape from causality violation for strong magnetic fields would require magnetic monopole/anti-monopole pair production, and this is relevant to Born's theory (among many others) because it is acausal for sufficiently strong magnetic fields even if the electric fields are small. 

Unfortunately, there is no known solution to the causality conditions \eqref{Schell1} and \eqref{Schell23} that could shortcut the task
of determining whether they are satisfied on a case by case basis. As we show here, however, there is a very simple way of implementing these conditions for the special subclass of  NLED theories that are electromagnetic-duality invariant. Following \cite{Ferko:2023wyi}, but also for brevity, we shall use the terminology ``self-dual'' for ``electromagnetic-duality invariant''.  
It should be understood that self-duality is never an  invariance of  $\mathcal{L}$, even for Maxwell.  It is an invariance of the Hamiltonian (and hence the field equations) under some special linear transformation of the Hamiltonian fields  $({\bf D},{\bf B})$ where ${\bf D}$ is the Legendre dual of ${\bf E}$.  Here, the term ``self-dual'' will indicate invariance of the Hamiltonian under an  $SO(2)$  `rotation' subgroup of $Sl(2;R)$. 

Self-dual theories include Maxwell and (as first observed by Schr\"odinger \cite{Schrodinger:1935oqa}) Born-Infeld electrodynamics. They also include the modified Maxwell (or ``ModMax'') theory with (source-free) Lagrangian density  \cite{Bandos:2020jsw,Bandos:2020hgy}
\begin{equation}\label{MM}
\mathcal{L} = (\cosh\gamma) S + (\sinh\gamma) \sqrt{S^2+P^2}\, , 
\end{equation}
where $\gamma\ge0$ is a dimensionless coupling constant ($\gamma<0$ is excluded by convexity, and Maxwell is recovered as the free-field $\gamma=0$ case). All self-dual NLED theories  that have a  weak-field limit  must reduce in this limit to ModMax, which we will take to include Maxwell, because this is the most general conformal self-dual  NLED with Lagrangian density $\LL (S,P)$. One non-conformal self-dual NLED with $\gamma>0$ is the  BI-type extension of ModMax  \cite{Bandos:2020jsw,Bandos:2020hgy}, which we called ``ModMaxBorn''  in \cite{Russo:2024kto}.  

Because self-duality is an invariance of the Hamiltonian, it is not manifest in a Lagrangian formulation. The condition on $\mathcal{L}$ required for self-duality was first found by Bialynicki-Birula  \cite{Bialynicki-Birula:1984daz} and has been rediscovered several times.  It takes the following simple form when $\mathcal{L}$ is viewed as a function of the Lorentz scalars 
$(U,V)$ \cite{Gibbons:1995cv}:
\begin{equation}\label{self-dual}
\mathcal{L}_U \mathcal{L}_V = -1\, . 
\end{equation}
The general solution to this equation has been known for a long time \cite{courant}, and was applied in a related context in \cite{Perry:1996mk}. It is given implicitly in terms of $(U,V)$ and an arbitrary real function $\ee (\tau)$ of  a real variable $\tau$:
\begin{equation}
\label{gensol}
\mathcal{L}=\ee (\tau)-\frac{2U}{[\dot {\ee }(\tau)]}\ ,\qquad 
\tau =V+\frac{U}{[\dot \ee (\tau)]^2}\, , 
\end{equation}
where $\dot\ee = d\ee /d\tau$. For the NLED application of this result, we see from \eqref{UVdef} that $\tau\geq 0$ with equality for 
$U=V=0$ (i.e. the vacuum). 

The formula \eqref{gensol} is the starting point for the results of this paper. Using it, the conditions \eqref{Schell1} and \eqref{Schell23} can be converted to conditions on the function $\ee$. Remarkably, because of the assumption of a weak-field limit, we find that the convexity conditions alone are both necessary and sufficient for causality, and that they are equivalent to the very simple inequalities  $\dot \ee\ge1$ and $\ddot \ee\ge0$. 

We then show how any NLED that is causal for weak fields but {\sl not} for sufficiently strong-fields determines a new causal, and self-dual, NLED with the same Lagrangian density at zero magnetic field. The function $\ell(\tau)$ for this `new' causal NLED is found from the Lagrangian density of the `old' acausal NLED at zero magnetic field.  A simple example is the Born theory, which is thus `converted' into Born-Infeld, but we use the example of  ``logarithmic electrodynamics'' to better illustrate the method.  

To begin, we take the exterior derivative on both sides of the two equations of \eqref{gensol} to get
\begin{equation}\label{dt}
\dot\ee ^2 d\mathcal{L} = Gd\tau -2\dot\ee \, dU \, , \quad Gd\tau = \dot\ee  ({\dot\ee }^2 dV + dU)\, , 
\end{equation}
where we have defined
\begin{equation}\label{Gee}
G := {\dot\ee }^3 + 2U\ddot\ee \, .  
\end{equation}
Combining these two equations we have
\begin{equation}
d\mathcal{L} = \dot\ee \,  dV - {\dot\ee }^{-1} dU\, , 
\end{equation}
which gives us 
 \begin{equation}\label{firstD}
\mathcal{L}_V=\dot\ee \, , \qquad \mathcal{L}_U=-1/\dot\ee \ . 
\end{equation}
Using this result, we verify that \eqref{self-dual} is solved by \eqref{gensol}, and we find that 
\begin{equation}
\sqrt{S^2+P^2}\ \mathcal{L}_S = V\mathcal{L}_V- U\mathcal{L}_U = \dot\ee  V + U/\dot\ee \, . 
\end{equation}
 As $(U,V)$ are both non-negative, the ``pre-condition'' \eqref{pre-c} (which, we recall, is a convention) is equivalent to 
\begin{equation}\label{dotg+}
\dot\ee >0\, . 
\end{equation}

Taking exterior derivatives again in \eqref{firstD}, and using the second equation of \eqref{dt} to eliminate $d\tau$, we find expressions for the second partial derivatives of $\mathcal{L}(U,V)$:
\begin{equation}\label{dderivs}
\mathcal{L}_{UU} =\ \frac{\ddot\ee }{\dot\ee  G}\ , \quad \mathcal{L}_{UV} =\ \frac{\dot\ee \ddot\ee }{G}\, , \quad
\mathcal{L}_{VV}  =\ \frac{\dot\ee ^3\ddot\ee }{G}\, . 
\end{equation}
This assumes that $G$ is nowhere zero. With this understood,  we now have expressions for both 
$\mathcal{L}_U$ and $\mathcal{L}_{UU}$ in terms of the first and second derivatives of $\ee $. Using them in \eqref{dos} yields the inequality $\dot\ee ^2/G >0$. As $\dot\ee >0$, this is equivalent to 
\begin{equation}\label{strong}
G>0\, , 
\end{equation}
which is therefore the strong-field causality condition for self-dual NLED. 
As we now explain, it has a simple geometric meaning. 

From \eqref{gensol}, we see that the curves of constant $\tau$ in the positive quadrant of the $(U,V)$-plane are straight lines that intersect both axes, $U=0$ and $V=0$. From the equation for $Gd\tau$ of \eqref{dt} it follows that if the lines of constant $\tau$ and $\tau +d\tau$ intersect then $G=0$ at the intersection. Thus, $G>0$ ensures that these lines never intersect, and therefore that they foliate the region of the positive $(U,V)$ quadrant in the domain of $\mathcal{L}(U,V)$. 

We now turn to the convexity conditions of \eqref{Schell1}. Although these become complicated when expressed 
in terms of second-derivatives of $\mathcal{L}(U,V)$,  they simplify for self-dual NLEDs when the equations \eqref{dderivs} 
for these second derivatives are used:
\begin{equation}\label{3Ldd}
\begin{aligned}
\mathcal{L}_{SS} &=\dot\ee ^{-1}\left[ A(\dot\ee ^2-1) + \mathcal{A} (\ddot\ee /G) \right] \, , \\
\mathcal{L}_{PP} &=\dot\ee ^{-1}\left[ B(\dot\ee ^2-1)+ \mathcal{B} (\ddot\ee /G)\right] \, , \\
\mathcal{L}_{SP} &= \dot\ee ^{-1}\left[ C(\dot\ee ^2-1)  + \mathcal{C} (\ddot\ee /G)\right]\, , 
\end{aligned}
\end{equation}
where 
\begin{eqnarray}\label{ABCs}
A &=& \frac{2UV}{(V+U)^3} \, , \qquad \mathcal{A} = \frac{(U- \dot\ee ^2 V)^2}{(V+U)^2}\, ,  \nonumber \\
B &=& \frac{(V-U)^2}{2(V+U)^3} \, , \qquad \mathcal{B} =\frac{(\dot\ee ^2+1)^2 UV}{ (V+U)^2}\, ,  \\
C &=& \frac{\sqrt{UV} (U-V)}{(V+U)^3}, \ \mathcal{C}=\frac{\sqrt{UV}(\dot\ee ^2 V-U)(\dot\ee ^2+1)}{(V+U)^2}.\nonumber
\end{eqnarray}
The third convexity inequality of \eqref{Schell1} now becomes
\begin{equation}
\frac{(U+V\dot\ee ^2)^2 }{2 (U+V)^3 \dot\ee ^2 }\ (\dot\ee ^2-1)\ \frac{\ddot\ee }{G}>0\ , 
\end{equation}
which requires  $(\dot\ee ^2-1)$ and $(\ddot\ee /G)$ to have the same sign. Since the factors $(A,\mathcal{A})$ and $(B,\mathcal{B})$ are non-negative, and generically positive, 
the first two convexity inequalities of \eqref{Schell1}  reduce (taking \eqref{dotg+} into account) to
\begin{equation}\label{first2}
\dot\ee  \ge1 \, , \quad \ddot\ee /G \ge0\, . 
\end{equation}

If these equations are combined with the strong-field causality constraint $G>0$ then
we arrive at the remarkably simple result that a self-dual NLED with a weak-field limit 
is causal if and only if
\begin{equation}\label{nands}
\dot\ee \ge1\, , \qquad \ddot\ee \ge0\, . 
\end{equation}
However, this result is already a consequence of the convexity constraints of \eqref{first2} because of the assumption of a weak-field limit. This assumption ensures  that  $|{\bf B}|=0$, and hence $U=0$, is part of the physical domain. 
%
At    $U=0$, one has $V=\tau $ and $G = \dot \ell ^3>0$,
and hence  $\ddot \ell \geq 0$ as a consequence of  the  second convexity  condition of  \eqref{first2}. 
This  holds for all $\tau $ for which $\ell(\tau ) $ is a real and non-singular function, and hence $G>0$ for any $U$ in the domain of $\mathcal{L}$.
In other words, {\sl weak-field causality implies strong-field causality for
any self-dual NLED with a weak-field limit}.

We shall now illustrate these results with some known self-dual NLED theories, and one new one:
\begin{itemize}

\item 
ModMax: $\ee (\tau)=a\tau$ for constant $a$ (the omission of a constant term corresponds to a choice of zero vacuum energy). 
This function yields 
\begin{equation}\
\mathcal{L}=  aV - a^{-1}U \, \qquad  \tau= V+ a^{-2}U\, . 
\end{equation}
The convexity condition $\dot\ee \ge1$ requires $a\ge1$, so we may set $a= e^\gamma$ for $\gamma\ge0$.  This yields
the ModMax Lagrangian density of \eqref{MM}, and Maxwell for $\gamma=0$.  
\end{itemize}
The fact that ModMax is the unique conformal weak-field limit of any non-conformal causal self-dual NLED can now be rephrased as a property of the function $\ee (\tau)$: it has a power series expansion about $\tau=0$ such that
\begin{equation}
\ee (\tau) = e^\gamma \tau + \mathcal{O}(\tau^2) \, . 
\end{equation}
An example for $\gamma=0$ is the BI theory, but this is the $\gamma=0$ 
case of our next example. 
\begin{itemize}
\item ModMaxBorn: $\ee (\tau) = T- \sqrt{T\left(T-2e^{\gamma}\tau\right)}$. Reality of $\ee $  requires 
$\tau\le e^{-\gamma} T/2$ and we now have $\dot\ee \ge1$ for $\gamma\ge0$, and $\ddot\ee >0$. All causality conditions are therefore satisfied for $\gamma\ge0$. 
The Lagrangian density is 
\begin{equation}
\label{MMB}
\quad  \mathcal{L} =  T - \sqrt{\left(T+ 2e^{-\gamma}U\right)\left(T -2e^\gamma V\right)} \, ,  
\end{equation}
which is BI for $\gamma=0$. For $\gamma>0$ it is the self-dual generalization of BI found in 
\cite{Bandos:2020jsw,Bandos:2020hgy}.  

\end{itemize}
Notice that an expansion of the ModMaxBorn function $\ee (\tau )$ in powers of $1/T$ is simultaneously a power-series expansion in $\tau$:
\begin{equation}
\ee (\tau) = e^\gamma \tau + \tfrac12 e^{2\gamma} \tau^2/T + \mathcal{O}(\tau^3)\, , 
\end{equation}
thus confirming that ModMax is the weak-field limit. Also, since $\tau=V$ at $U=0$, the maximum of 
$\tau$ allowed by reality of $\ee(\tau)$ is also the maximum value of $V$ allowed for reality of the ModMaxBorn Lagrangian density. 

Recall that the curves of constant $\tau$ in the positive quadrant of the $(U,V)$-plane are always straight lines, and for causal theories they foliate the region of this quadrant in the domain of 
$\mathcal{L}(U,V)$. From the above examples we see that the lines are parallel for ModMax and that they foliate the entire quadrant. More generally the magnitude of the slope of the lines decreases with increasing $\tau$ (since $\ddot\ee >0$) and the lines are never parallel. There are then two possibilities. If the slope becomes zero at a $\tau=\tau_{\rm max}$ then $V(\tau_{\rm max})=V_{\rm max}$ and the lines of constant $\tau$ foliate the $V<V_{\rm max}$  subspace of the quadrant; this possibility is illustrated by ModMaxBorn. The other possibility is that there is no maximum of $\tau$, or $V$, and the lines of constant $\tau$ foliate the entire positive quadrant; this is illustrated by the following new example: 
\begin{itemize}
\item $\ee (\tau)= \frac23 e^\gamma T(1+ \tau/T)^\frac32$. We have $\ddot\ee >0$ for all $\tau$ but the condition $\dot\ee >1$ requires $\gamma\ge0$, so the theory is causal for $\gamma\ge0$. The power-series expansion of $\ee (\tau)$ shows that the weak-field limit is ModMax, but in contrast to ModMaxBorn there is no maximum value of $\tau$.  
The Lagrangian density is  
\begin{equation}
\ \ \mathcal{L}=\frac23 e^\gamma T \left\{\sqrt{2}\left(1+\frac{V}{T}-\frac{\Delta}{2} \right) \sqrt{1+\frac{V}{T}+\Delta}\right\}\, , 
\end{equation}
where $\Delta= \sqrt{\left(1+\frac{V}{T}\right)^2 + 4e^{-2\gamma} \frac{U}{T}}$. 
Notice that reality of $\mathcal{L}$ imposes no constraint on either $U$ or $V$. 

\end{itemize}

We have now seen how the Lagrangian density $\mathcal{L}$ for a self-dual NLED may be found from its associated one-variable function $\ee(\tau)$, but how do we find the function $\ee(\tau)$ given $\mathcal{L}$? To answer this question, we first 
observe that $\mathcal{L}$ is a function only of $E=|{\bf E}|$ when $|{\bf B}|=0$; call this function $L(E)$. Next, we observe that 
$(U,V)=(0, \frac12 E^2)$ for zero magnetic field, and hence, from \eqref{gensol}, 
\begin{equation}
\mathcal{L}(E) = \ee(\tau)\, , \quad \tau= \frac12 E^2 \qquad (|{\bf B}|=0). 
\end{equation}
In other words, $\ee(\tau)$ is the function found from the Lagrangian density at $|{\bf B}|=0$ by setting $E= \sqrt{2\tau}$ (as is easily verified for the above examples). 

This suggests a categorisation of NLED theories according to the function $L(E)$ that they define at
$|{\bf B}|=0$. For each such function there is an infinite number of possible Lagrangian densities
$\mathcal{L}(S,P)$ but only one is self-dual. We can exploit this observation to `convert' any non-self-dual NLED into the self-dual NLED in its category, which will be causal if the associated function $\ee(\tau)$ satisfies the conditions of \eqref{nands}. 

This construction is well-illustrated by the ``logarithmic electrodynamics'' of \cite{Soleng:1995kn}, which (like all variants of it analysed in \cite{Russo:2024kto}) is acausal for strong fields. However,
the self-dual version is causal; its construction below allows for its $\gamma>0$ generalisation:
\begin{itemize}

\item $\ee (\tau)=- e^\gamma T \ln(1-\tau/T)$. This is defined for $\tau<T$, and the causality conditions \eqref{nands}  are satisfied in this range if $\gamma\ge0$. The power-series 
expansion of $\ee (\tau)$ shows that ModMax (Maxwell for $\gamma=0$) is the weak-field limit. The equation for $\tau$ is quadratic in this case but has only one positive solution, which yields 
\begin{equation}
    \mathcal{L} = - e^\gamma T\ln (\Xi/U) -2 e^{-\gamma}\Xi\ ,
\end{equation}
where 
\begin{equation}
    \Xi= \frac12 e^{2\gamma}\left[\sqrt{T^2 + 4e^{-2\gamma}(T-V)U} -T \right]\ .
 \end{equation}
 The special (acausal) case with $\gamma =-\ln(2)$ was found previously in 
 \cite{Mkrtchyan:2022ulc} using a different method. 
\end{itemize}

We conclude with the observation that the Lagrangian density for general {\sl causal} self-dual NLED can be written as 
\begin{equation}
\label{Laux}
    \mathcal{L} = \ee (\tau) -\frac{2U}{\dot \ee (\tau)} -  \lambda \left(\tau - V - \frac{U}{[\dot \ee (\tau)]^2}\right)\, , 
\end{equation} 
where $\lambda$ is a Lagrange multiplier field. 
For a causal self-dual NLED theory the fields $(\lambda,\tau)$ are an auxiliary pair because the field equation of one yields an equation that uniquely determines the other, as we now explain.

Variation of $\tau$ in \eqref{Laux} yields the equation  $G[\lambda - \dot \ee (\tau)]=0$, which is equivalent to $\lambda= \dot\ee(\tau)$ only if $G\ne0$ as is the case for causal theories, since causality requires $G>0$ (in the domain 
of $\LL$). Similarly, variation of $\lambda$ in \eqref{Laux} yields the equation for $\tau$ of \eqref{gensol}, which uniquely determines $\tau$ as a function of $(U,V)$ as long as $G\ne0$. This is because the only way that a point in the positive $(U,V)$ quadrant can fail to have a definite value of $\tau$ is if it is an intersection point of two (or more) distinct lines of constant $\tau$, but this can happen only at points for which $G=0$. The pair of fields $(\lambda,\tau)$ is therefore a {\it bona fide} auxiliary-field pair for all causal self-dual NLED theories, and their elimination yields the Lagrangian density defined by \eqref{gensol}. 

In contrast to previously proposed auxiliary-field formulations of generic self-dual NLED theories (e.g. \cite{Ivanov:2003uj,Avetisyan:2021heg}) the implementation 
of causality in \eqref{Laux} is very simple: one has only to choose the function $\ell$ such  that the conditions 
\eqref{nands} are satisfied. In addition, \eqref{Laux} provides a natural starting point for a generalisation of causal self-dual NLED theories to arbitrary spacetimes and hence for the inclusion of gravity, but we leave this for future investigations.


\smallskip
\noindent\textbf{Acknowledgements}: We thank Dima Sorokin for pointing out an error in the previous version of this paper. PKT has been partially supported by STFC consolidated grant ST/T000694/1. JGR acknowledges financial support from a MINECO grant PID2019-105614GB-C21.

\end{document}